\documentclass[conference]{IEEEtran}
\IEEEoverridecommandlockouts
\usepackage{cite}
\usepackage{amsmath,amssymb,amsfonts}
\usepackage[linesnumbered,ruled,vlined]{algorithm2e}  
\usepackage{multirow}  
\usepackage{hyperref}  
\usepackage{fancyhdr}
\usepackage{float}  
\usepackage{tabulary}  
\usepackage{graphicx}
\usepackage{textcomp}
\usepackage[table]{xcolor}
\def\BibTeX{{\rm B\kern-.05em{\sc i\kern-.025em b}\kern-.08em
    T\kern-.1667em\lower.7ex\hbox{E}\kern-.125emX}}
\begin{document}

\title{Exact Distributed Stochastic Block Partitioning\\
\thanks{DISTRIBUTION STATEMENT A. Approved for public release. Distribution is unlimited. This material is based upon work supported by the Under Secretary of Defense for Research and Engineering under Air Force Contract No. FA8702-15-D-0001. Any opinions, findings, conclusions or recommendations expressed in this material are those of the author(s) and do not necessarily reflect the views of the Under Secretary of Defense for Research and Engineering. \copyright 2022 Massachusetts Institute of Technology. Delivered to the U.S. Government with Unlimited Rights, as defined in DFARS Part 252.227-7013 or 7014 (Feb 2014). Notwithstanding any copyright notice, U.S. Government rights in this work are defined by DFARS 252.227-7013 or DFARS 252.227-7014 as detailed above. Use of this work other than as specifically authorized by the U.S. Government may violate any copyrights that exist in this work.}
}

\author{\IEEEauthorblockN{Frank Wanye}
\IEEEauthorblockA{\textit{Dept. of Computer Science} \\
\textit{Virginia Tech}\\
Blacksburg, VA, USA \\
wanyef@vt.edu}
\and
\IEEEauthorblockN{Vitaliy Gleyzer}
\IEEEauthorblockA{\textit{MIT Lincoln Laboratory} \\
Lexington, MA, USA \\
vgleyzer@ll.mit.edu}
\and
\IEEEauthorblockN{Edward Kao}
\IEEEauthorblockA{\textit{MIT Lincoln Laboratory} \\
Lexington, MA, USA \\
edward.kao@ll.mit.edu}
\and
\IEEEauthorblockN{Wu-chun Feng}
\IEEEauthorblockA{\textit{Dept. of Computer Science} \\
\textit{Virginia Tech}\\
Blacksburg, VA, USA \\
wfeng@vt.edu}}

\maketitle

\pagestyle{fancy}
\fancyhf{}
\fancyhead[L]{This work has been submitted to the IEEE for possible publication. Copyright may be transferred without notice, after which this version may no longer be accessible.}
\fancyfoot[R]{\thepage}
\thispagestyle{fancy}

\begin{abstract}
Stochastic block partitioning (SBP) is a community detection algorithm that is highly accurate even on graphs with a complex community structure, but 
its inherently serial nature hinders its widespread adoption by the wider scientific community.
To make it practical to analyze large real-world graphs with SBP, there is a growing need to parallelize and distribute the algorithm. The current state-of-the-art distributed SBP algorithm is a divide-and-conquer approach that limits communication between compute nodes until the end of inference. 
This leads to the breaking of computational dependencies, which causes convergence issues as the number of compute nodes increases, and when the graph is sufficiently sparse.
In this paper, we introduce EDiSt - an \underline{e}xact \underline{di}stributed \underline{st}ochastic block partitioning algorithm. Under EDiSt, compute nodes periodically share community assignments during inference. Due to this additional communication, EDiSt improves upon the divide-and-conquer algorithm by allowing it to scale out to a larger number of compute nodes without suffering from convergence issues, even on sparse graphs. We show that EDiSt provides speedups of up to 23.8$\times$ over the divide-and-conquer approach, and speedups up to 38.0$\times$ over shared memory parallel SBP when scaled out to 64 compute nodes.
\end{abstract}


\section{Introduction}








Much of the data that is collected today contains entities that are connected to each other. For example, in network traffic data, network nodes communicate with each other via packets, and in social media data, people interact with each other's posts and add each other to contact lists. Such data can be represented in the form of a graph, where the entities are represented by a graph's vertices, and the relationships between the entities are represented by the edges between vertices. Often, these vertices form structural groups, called communities, such that the vertices within each group are more closely connected to each other than they are to vertices in other groups. The process of identifying these groups is referred to as community detection~\cite{Fortunato2010CommunityGraphs}.

Due to the expressive nature of the graph representation, and the ubiquity of community structure in real-world graph data, community detection has a lot of applications across a wide variety of fields. These fields include bioinformatics, where community detection has been used to aid in identifying carcinogenic gene combinations~\cite{Oles2023BiGPICC:Data}, communication networks, where community detection can be used to aid in the placement of servers~\cite{Krishnamurthy2000OnClients}, and artificial intelligence, where community detection can be used as a pre-processing step that leads to improved classification accuracy~\cite{Rizos2017MultilabelNetworks,Li2017PacketDetection}. The proliferation of these use cases has led to a lot of interest in research involving community detection.

Optimal community detection is an NP-hard problem. Hence, various heuristics are used to approximate the community structure of a graph~\cite{Fortunato2010CommunityGraphs}. One such heuristic is stochastic blockmodel inference~\cite{Peixoto2013ParsimoniousNetworks}. This heuristic involves building a blockmodel, a latent model of the inter-community connectivity in a graph~\cite{Karrer2011StochasticNetworks}, and then iteratively perturbing this model to find the combination of number of communities and vertex-to-community assignments that minimize the description length of the blockmodel.

Stochastic block partitioning~(SBP)~\cite{Peixoto2013ParsimoniousNetworks,Peixoto2014EfficientModels} is a community detection algorithm based on stochastic blockmodel inference. 
 It is of interest to the research community due to its high accuracy, even when the graph community structure is complicated by highly varied community sizes, varied degree distributions within communities, and a high degree of inter-community connectivity, as can be seen by comparing the results obtained on the IEEE/Amazon/MIT Graph Challenge~\cite{Kao2017StreamingPartition}. However, SBP is slower than competing community detection algorithms based on other heuristics, and difficult to parallelize owing to the fact that it is based on an inherently sequential Markov chain Monte Carlo algorithm~\cite{Kao2017StreamingPartition}.

There have been several attempts to accelerate SBP through sampling~\cite{Wanye2019FastSampling}, shared memory parallelism~\cite{Wanye2022OnDetection}, and various optimizations~\cite{Uppal2021FasterControl}. However, 
for the runtime of SBP to be practical on large-scale, real-world graphs, which often consist of upwards of millions of vertices and edges, the computation needs to be distributed across multiple compute nodes.

To the best of our knowledge, only one distributed implementation of stochastic block partitioning exists - the divide-and-conquer approach described in~\cite{Uppal2017ScalablePartition}, henceforth abbreviated as DC-SBP. DC-SBP works in a similar fashion to the Map-Reduce paradigm, with no communication between compute nodes until the "Reduce" phase, where partial results are combined. Our testing of DC-SBP reveals that there are two fairly common conditions under which the approach suffers from convergence issues that a) lead to severe degradation in the quality of community structure that it infers and b) limit the number of compute nodes that DC-SBP can scale to while maintaining accuracy (see Section~\ref{sec:dcsbpablationresults}). In this work, we introduce a novel \underline{e}xact \underline{di}stributed \underline{st}ochastic block partitioning~(EDiSt) algorithm that does not suffer from these convergence issues owing to the introduction of periodic inter-node MPI communication during the course of inference. Thus, EDiSt enables distributed SBP on larger computational clusters, and a larger selection of graphs without sacrificing result quality. 
The differences between DC-SBP and EDiSt are summarized in Table~\ref{tab:introdifferences}.


\begin{table}[htbp!]
\caption{Differences between EDiSt and DC-SBP}
\begin{center}
\begin{tabulary}{\columnwidth}{|L|L|L|}
\hline
Parameter & DC-SBP & EDiSt \\
\hline
Data handling & Round-robin data distribution & Data duplication \\
Inter-node communication & Limited to partial result re-combination & Periodic throughout inference \\
Retains accuracy beyond 16 MPI ranks & No & Yes \\
Retains accuracy on sparse graphs & No & Yes \\
\hline
\end{tabulary}
\label{tab:introdifferences}
\end{center}
\end{table}

Our main contributions can be summarized as follows:

\begin{itemize}
    \item We empirically demonstrate two conditions under which divide-and-conquer distributed stochastic block partitioning suffers from convergence issues.
    \item We introduce EDiSt, a novel distributed stochastic partitioning algorithm. We empirically show that EDiSt does not suffer from convergence issues under the same conditions as the divide-and-conquer approach, which allows it to make use of more computational nodes and run on a larger selection of graphs without sacrificing result quality.
    \item We evaluate the scalability of EDiSt, showing that by scaling it out to 64 compute nodes on a compute cluster, we can not only achieve up to a 38.0$\times$ speedup over shared memory parallel SBP, but also a 23.8$\times$ speedup over the divide-and-conquer distributed SBP approach.
\end{itemize}

\section{Background and Related Work}

In this section, we provide some background on stochastic blockmodels, the stochastic block partitioning algorithm, and the state-of-the-art in distributed stochastic block partitioning.

\subsection{Stochastic Blockmodels}

Stochastic blockmodels~(SBMs)~\cite{Karrer2011StochasticNetworks} are a class of generative models that describe the structure of a graph based on its community structure. They can be used to either generate graphs with a specified community structure, or to infer the community structure of a specified graph.

Typically, a stochastic blockmodel is represented by a matrix $M$ of size $C \times C$, where $C$ refers to the number of communities present in the model. Every entry $M_{a,b}$ of the blockmodel matrix corresponds to the number of edges between (or, if the blockmodel is not microcanonical, the probability of edges forming between), two communities $a$ and $b$.

A quality function is used to fit blockmodels to a given graph when using them to perform community detection. One such function is the log-likelihood of the graph $G$ given the blockmodel $B$, $L(G|B)$. The log-likelihood function varies between the different variants of SBMs. In this paper, we focus on the degree-corrected SBM~(DCSBM), which accounts for differences in degree distribution between communities in a graph. The log-likelihood for the DCSBM is given by the following equation~\cite{Kao2017StreamingPartition,Peixoto2013ParsimoniousNetworks}:

\begin{equation}\label{eq:dcsbm_log_likelihood}
    L(G|B) = \sum_{i,j}{B_{i,j}\log{\left(\frac{B_{i,j}}{d_i^{out}  d_j^{in}}\right)}},
\end{equation}

where $B_{i,j}$ is the number of edges between communities $i$ and $j$, and $d_i^{out}$ and $d_j^{in}$ are the out-degree of community $i$ and the in-degree of community $j$, respectively, for a directed graph.

However, the log-likelihood function is unsuitable for community detection where the number of communities is not known apriori. This is because $L(G|B)$ will be maximized when the number of communities $C$ is equal to the number of vertices $V$. To overcome this limitation, when the optimal number of communities needs to be inferred alongside the optimal vertex-to-community assignment, the description length metric, $DL$, is used instead of $L(G|B)$. For a directed graph, $DL$ is given by the following equation~\cite{Kao2017StreamingPartition,Peixoto2012EntropyEnsembles}:

\begin{equation}\label{eq:dcsbm_dl}
    DL = Eh\left(\frac{C^2}{E}\right) + V\log{C} - L(G|B),
\end{equation}

where $E$ is the number of edges in the graph, $C$ is the number of communities in the graph, $h(x) = (1 + x)\log{1 + x} - x\log{x}$, $V$ is the number of vertices in the graph and $L(G|B)$ is given by Equation~\ref{eq:dcsbm_log_likelihood}. Unlike $L(G|B)$, $DL$ is minimized when used to infer the community structure of a graph.

\subsection{Stochastic Block Partitioning}

Stochastic block partitioning~(SBP)~\cite{Peixoto2013ParsimoniousNetworks,Peixoto2014EfficientModels,Kao2017StreamingPartition} is a community detection algorithm based on inference over the degree-corrected stochastic blockmodel~(DCSBM). It works by minimizing the description length of the DCSBM~(see Equation~\ref{eq:dcsbm_dl}) using Markov chain Monte Carlo~(MCMC) techniques.

\begin{figure*}[htbp!]
  \centering
  \includegraphics[width=0.7\textwidth]{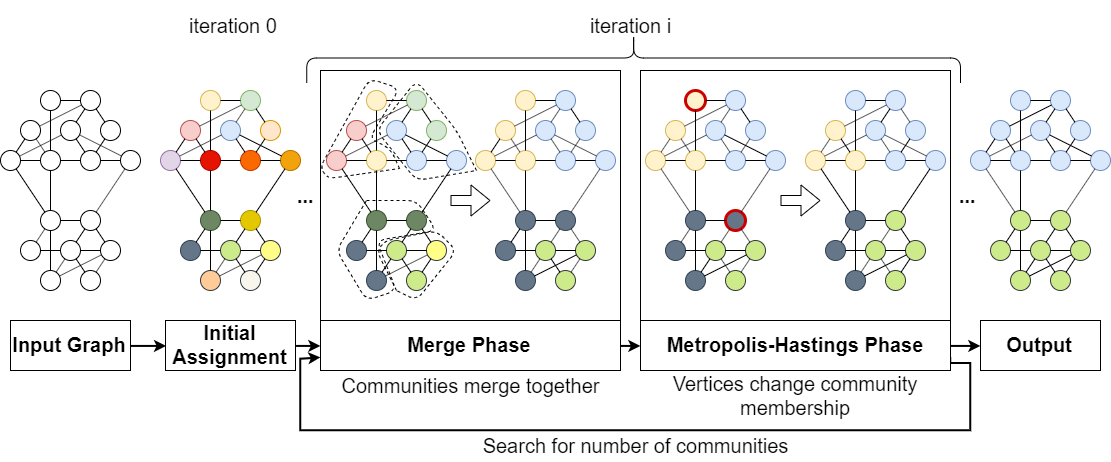}
  \caption{Snapshots of a graph at various stages of the stochastic block partitioning algorithm.}
  \label{fig:sbp}
\end{figure*}

This optimization is iterative and agglomerative, and is executed in two alternating phases -- the block merge phase and the MCMC phase. In the MCMC phase, described in Algorithm~\ref{alg:mcmc_phase}, the inherently sequential Metropolis-Hastings algorithm~\cite{Hastings1970MonteApplications} is used to move individual vertices from one community to another, based on the change in the description length of the DCSBM. In the block merge phase, described in Algorithm~\ref{alg:block_merge}, entire communities~(or blocks) are merged together, 
reducing the likelihood of the MCMC process getting stuck in a local minimum. The algorithm's execution is summarized in Fig.~\ref{fig:sbp}.

In order to allow the algorithm to automatically find the optimal number of communities, a golden ratio search is used. In this search algorithm, up to 3 versions of the blockmodel are stored in decreasing order of number of communities. So long as all 3 blockmodels are also in decreasing order of description length, the next phase of the algorithm will start with the blockmodel with the smallest number of communities and proceed with the block merge phase. If however, the subsequent blockmodel results in an increased description length, then the golden ratio criterion is met, and the optimal number of communities is within the range specified by the 3 stored blockmodels.

A parallel formulation of SBP was proposed in~\cite{Wanye2022OnDetection}. This formulation replaced the Metropolis-Hastings portion of the algorithm with a hybrid algorithm that processes informative, high-degree vertices sequentially, and the less informative low-degree vertices in parallel via an adaptation of asynchronous Gibbs sampling~\cite{Terenin2018}.


\begin{algorithm}[htbp]
    \SetKwFor{RepTimes}{repeat}{times}{end}
    \SetKwFor{ForBy}{for}{do in parallel}{end for}
    \SetAlgoLined
    \KwData{Graph $G$, Blockmodel $B$, int $x$}
    \KwResult{Updated Blockmodel $B$}
    initialize best\_merges container\;
    \ForBy{community $c \in B$}{
        \RepTimes{$x$}{
            Propose a new community $c'$ to merge with $c$\;
            Calculate $\Delta DL$ when $c$ is merged with $c'$\;
            \If{$\Delta DL$ is best obtained so far for $c$}{
                Store $c'$ and $\Delta DL$ for $c$ in best\_merges\;
            }
        }
    }
    sort best\_merges on $\Delta DL$\;
    \Repeat{number of communities is halved}{
        $c$, $c'$ = best\_merges.pop()\;
        Merge $c$ into $c'$\;
    }
\caption{Block Merge Phase}
\label{alg:block_merge}
\end{algorithm}

\begin{algorithm}[htbp]
    \SetKwFor{RepTimes}{repeat}{times}{end}
    \SetAlgoLined
    \KwData{Graph $G$, Blockmodel $B$, double $t$, int $x$}
    \KwResult{Updated Blockmodel $B$}
    compute MDL of $B$\;
    \Repeat{$\Delta DL < t \times DL$ or $x$ times}{
        \ForEach{vertex $v \in G$}{
            propose new community $c$ for $v$\;
            compute $\Delta DL$ for proposed move\;
            compute Metropolis-Hastings ratio from $\Delta DL$\;
            \If{move is accepted}{
                move $v$ to $c$ and update $B$\;
            }
        }
        compute MDL of $B$\;
    }
\caption{MCMC Phase}
\label{alg:mcmc_phase}
\end{algorithm}


\subsection{Divide-and-Conquer SBP}

The data access patterns of SBP make distributing the algorithm a non-trivial task. For every proposed change in community membership, the algorithm needs access to at least two rows \textit{and} two columns of the SBM matrix. Thus, a row- or column-wise distribution of the blockmodel would lead to all-to-one communication for every proposed change in community membership. Moreover, any accepted changes in community membership (and the corresponding changes to the blockmodel rows and columns) would need to be broadcast to every other worker. A ``traditional'' distribution of the algorithm would therefore be ineffective due to excessive amounts of communication.

To the best of our knowledge, the only published distributed SBP algorithm is the MPI-based divide-and-conquer (DC-SBP) algorithm of Uppal, Swope and Huang~\cite{Uppal2017ScalablePartition}, developed as part of the MIT/Amazon/IEEE Graph Challenge~\cite{Kao2017StreamingPartition}. DC-SBP divides the graph into $n$ subgraphs, where $s$ is the number of MPI ranks/tasks. Each rank then runs the SBP algorithm on its subgraph independently, until the golden ratio criterion is reached. The resulting community memberships are then sent to the root rank, where they are combined. This combination happens in two steps.

In the first step, community memberships from successive subgraph pairs are iteratively combined, such that each 
community from one subgraph is merged into a community from the other subgraph. The merges are selected based on the best potential change in DL. This is repeated until the number of subgraphs is reduced to a threshold $t$, which was chosen as 4 by the authors. This allows the algorithm to potentially recover if some subgraphs do not contain the same communities.

In the second step, the remaining community memberships are merged to form the community memberships for the entire graph. 
SBP then continues on the whole graph using the root rank, thus fine-tuning the merged results and allowing the algorithm to find the optimal number of communities.

The pseudocode for DC-SBP is given in Alg.~\ref{alg:dcsbp}. The authors have made their code available at \url{https://github.com/iHeartGraph/GraphChallenge}. It is developed in python using NumPy~\cite{vanderWalt2011TheComputation} for array operations and the mpi4py library~\cite{Dalcin2008MPIExtensions} for distributed computation. Since python is an interpreted language, this python implementation can be significantly sped up by translating it to C++.

\begin{algorithm}[h]
    \SetKwFor{RepTimes}{repeat}{times}{end}
    \SetAlgoLined
    \KwData{Graph $G$, MPI rank $r$}
    \KwResult{Blockmodel $B$}
    subgraph $G^r$ = round robin sample from $G$ based on $r$\;
    partial blockmodel $B^r$ = create initial blockmodel from $G^r$\;
    $B^r$ = SBP($G^r$, $B^r$)\;
    partial results $R$ = list($B^r$)\;
    \eIf{$r == 0$}{
        \ForEach{rank $r', r' \neq 0$}{
            partial result $B^{r'}$ = MPI\_Recv($r'$)\;
            append $B^{r'}$ to $R$
        }
    }{
        MPI\_Send $B^r$ to rank 0\;
        return\;
    }
    \Repeat{$R.length \leq 4$}{
        $R'$ = list()\;
        \ForEach{successive $B^{r1}, B^{r2} \in R$}{
            $B'$ = merge $B^{r2}$ communities into $B^{r1}$ communities\;
            append $B'$ to $R'$
        }
        $R = R'$\;
    }
    intermediate blockmodel $B'$ = merge partial results in $R$\;
    return SBP($B'$)\;
\caption{Divide-and-Conquer SBP}
\label{alg:dcsbp}
\end{algorithm}


The disconnectedness between the subgraphs in DC-SBP results in minimal communication between MPI ranks/tasks, which theoretically allows the algorithm to efficiently scale to a large number of nodes, and neatly solves the problem of how to distribute the blockmodel matrix. However, it also creates to convergence issues by breaking several computational dependencies in the graph. Therefore, in practice, scaling the algorithm to a large number of MPI tasks leads to severe accuracy degradation~(see Section\ref{sec:dcsbpablationresults}). In~\cite{Uppal2017ScalablePartition}, the authors report achieving good results with 8 MPI tasks.

Additionally, the fine-tuning phase happens on a single computational node. These fine-tuning SBP iterations are faster than the initial ones due to a greatly decreased number of communities (assuming that the number of communities is a small fraction of the number of vertices, which is usually true in practice), and a decrease in the number of vertices that change community membership, which leads to fewer blockmodel updates being performed. However, on large graphs, they can still present a significant performance bottleneck.






\section{Method}

In this section, we describe our baseline DC-SBP implementation, as well as our exact distributed stochastic block partitioning algorithm.

\subsection{C++ Divide-and-Conquer SBP Implementation}

In order to speed up the Divide-and-Conquer SBP implementation, we first translate the python implementation to C++, using MPI for communication between nodes. We then optimize the translated implementation for faster execution on sparse graphs. Some of the optimizations implemented include a) the use of a sparse matrix in the form of a vector of hashmap objects to store the blockmodel matrix, b) storing the transpose of the blockmodel matrix for fast access along both rows and columns, c) using a sparse vector of changes to the blockmodel to perform change in description length computations, and d) using a pointer-based scheme to keep track of the community merges during the block merge phase of SBP, which speeds up the block merge phase of the algorithm.


To parallelize the computation within an MPI rank, we use the Hybrid SBP~\cite{Wanye2022OnDetection} algorithm implemented using OpenMP.

\subsection{Exact Distributed SBP}

The main difference between the DC-SBP algorithm and our exact distributed approach lies in the inter-subgraph connectivity. In DC-SBP, the subgraphs are disconnected, and an MPI rank working on subgraph A has no knowledge of the vertices or the communities in subgraph B. As such, the more ranks (and subgraphs) there are, the less information a rank has access to, and the higher the chances that the SBP algorithm fails to converge. The upside of this lack of connectivity is that processing each subgraph requires much less memory, and is much faster, both due to the super-linear runtime of SBP and the minimal inter-node communication.

In our exact distributed approach, we allow each rank to have knowledge of the vertices and communities present in the other ranks. Each rank stores the blockmodel for the entire graph, but is responsible for the computation of only a portion of the communities or vertices, depending on the phase of the algorithm. Since communicating every change to the blockmodel is impractical due to the communication burden this would impose, we allow the ranks to sync their blockmodels at the end of every block merge phase, and after each MCMC iteration.

In the block merge phase, each rank is responsible for computing the merge proposals and corresponding changes in description length for a disjoint set of communities. Once these computations are done, the ranks communicate the merge proposal with the best (most negative) change in description length for each of the communities they are responsible for, to all the other ranks using MPI all-to-all communication primitives. Each rank then performs the requisite number of merges by selecting the best merges from all the ranks, and updates its blockmodel accordingly. The pseudocode for this phase is shown in Algorithm~\ref{alg:dist_block_merge}.

In the MCMC phase, each rank is responsible for computing change in community membership proposals and the corresponding changes in description length for a disjoint set of vertices. Each rank makes a pass over the vertices assigned to it, updating its local copy of the blockmodel as needed, and storing the move in a vector. At the end of the pass, the ranks exchange the accepted moves using MPI all-to-all communication primitives. Then, each rank updates their local copy of the blockmodel by performing the moves received from all other ranks. The pseudocode for this phase is shown in Algorithm~\ref{alg:dist_mcmc_phase}. Note that for the sake of consistency, the pseudocode shown is for the sequential, Metropolis-Hastings-based MCMC phase. However, the differences between the distributed Hybrid MCMC phase used in this work and the distributed Metropolis-Hastings MCMC phase are minimal.

\begin{algorithm}[htbp]
    \SetKwFor{RepTimes}{repeat}{times}{end}
    \SetKwFor{ForBy}{for}{do in parallel}{end for}
    \SetAlgoLined
    \KwData{Graph $G$, Blockmodel $B$, int $x$, MPI rank $r$}
    \KwResult{Updated Blockmodel $B$}
    N = MPI\_Comm\_size(MPI\_COMM\_WORLD)\;
    initialize best\_merges container\;
    \ForBy{community $c \in B$}{
        \If{$c \;\mathrm{mod}\; N \neq r$}{
            continue\;
        }
        \RepTimes{$x$}{
            Propose a new community $c'$ to merge with $c$\;
            Calculate $\Delta DL$ when $c$ is merged with $c'$\;
            \If{$\Delta DL$ is best obtained so far for $c$}{
                Store $c'$ and $\Delta DL$ for $c$ in best\_merges\;
            }
        }
    }
    MPI\_Allgather(best\_merges)\;
    sort best\_merges on $\Delta DL$\;
    \Repeat{number of communities is halved}{
        $c$, $c'$ = best\_merges.pop()\;
        Merge $c$ into $c'$\;
    }
\caption{Block Merge Phase in EDIST}
\label{alg:dist_block_merge}
\end{algorithm}

\begin{algorithm}[htbp]
    \SetKwFor{RepTimes}{repeat}{times}{end}
    \SetAlgoLined
    \KwData{Graph $G$, Blockmodel $B$, double $t$, int $x$, list $V'$ of vertices this rank is responsible for}
    \KwResult{Updated Blockmodel $B$}
    compute MDL of $B$\;
    \Repeat{$\Delta DL < t \times DL$ or $x$ times}{
        initialize accepted\_merges container\;
        \ForEach{vertex $v \in G$}{
            \If{$v \notin V'$}{
                continue\;
            }
            propose new community $c$ for $v$\;
            compute $\Delta DL$ for proposed move\;
            compute Metropolis-Hastings ratio from $\Delta DL$\;
            \If{move is accepted}{
                move $v$ to $c$ and update $B$\;
                store $v$ and $c$ in accepted\_merges\;
            }
        }
        MPI\_Allgather(accepted\_merges)\;
        \ForEach{$v$ and $c$ pair in accepted\_merges}{
            \If{vertex $v$ is not in community $c$}{
                move $v$ to $c$ and update $B$\;
            }
        }
        compute MDL of $B$\;
    }
\caption{MCMC Phase in EDIST}
\label{alg:dist_mcmc_phase}
\end{algorithm}

In order to maintain load balance in the MCMC phase, we adopt an approach similar to the sorting-based blocking algorithm described in~\cite{Yu2019GPU-BasedReconstructions}. We first sort the vertices according to vertex degree. Then, assuming there are $n$ ranks, rank $r$ gets assigned the set of vertices $(r, 2n - r, 2n + r, 4n - r, 4n + r, 6n - r, ...)$. This has the effect of breaking the sorted set of vertices in chunks of $2n$, and then assigning to each rank the vertices with the $r$th highest and $r$th lowest degrees.

The syncing of the blockmodels between ranks, and the fact that each rank keeps track of the entire blockmodel, mean that the exact distributed SBP approach is likely to be slower and require more memory per rank than DC-SBP. At the same time, these weaknesses allow it to scale to a much higher number of ranks before accuracy degradation can be expected to set in. On graphs with a particularly sparse dependency distribution, it is theoretically possible to scale the approach to $V$ MPI ranks, where $V$ is the number of vertices, without any accuracy degradation.

\section{Experimental Setup}

In this section, we describe the datasets and hardware used to evaluate DC-SBP and EDiSt.

\subsection{Synthetic Datasets}

In order to compare the results of our approach with those presented in~\cite{Uppal2017ScalablePartition}, we use the datasets published in the MIT/Amazon/IEEE Graph Challenge~\cite{Kao2017StreamingPartition}. These graphs were generated using the `graph-tool`~\cite{Peixoto2014TheLibrary} python library, by generating a blockmodel with the desired characteristics, and then perturbing a random graph until it matches these characteristics as closely as possible. The selected graphs are summarized in Table~\ref{graph_challenge_datasets}. The graphs labeled as `easy` correspond to those with a low block overlap and low block size variation, and those labeled as `hard` correspond to those with a high block overlap and high block size variation.

\begin{table*}[!htbp]
\caption{Graph Challenge Datasets}
\begin{center}
\begin{tabular}{|c|r|r|c|c|c|}
\hline
ID & Number of & Number of & Community & Community Size & Number of \\
& Vertices & Edges & Overlap & Variation & Communities \\
\hline
20K-easy & $20,000$ & $473914$ & low & low & $32$ \\
20k-hard & $20,000$ & $473329$ & high & high & $32$ \\
50K-easy & $50,000$ & $1183975$ & low & low & $44$ \\
50k-hard & $50,000$ & $1187682$ & high & high & $44$ \\
200k-easy & $200,000$ & $4750333$ & low & low & $71$ \\
200k-hard & $200,000$ & $4754406$ & high & high & $71$ \\
\hline
\end{tabular}
\label{graph_challenge_datasets}
\end{center}
\end{table*}

When testing the divide-and-conquer approach, we noticed significant differences in how well the algorithm retains accuracy as the number of nodes increases based on the graph structure. On the graph challenge~\cite{Kao2017StreamingPartition} graphs, the divide-and-conquer approach maintained accuracy until between 8 and 16 MPI ranks. But on the web-graph-like graphs used in~\cite{Wanye2022OnDetection}, the divide-and-conquer approach sometimes failed to maintain accuracy even with 2 MPI ranks. This is despite the fact that both sets of graphs were generated using a similar approach, using the blockmodel-based generator from the `graph-tool`~\cite{Peixoto2014TheLibrary} python library. In order to determine the cause of this difference, we identified the following differences between the two sets of graphs:

\begin{itemize}
    \item Number of communities: the graph challenge graphs had a significantly smaller ratio of communities to vertices.
    \item Degree sequence duplication: the `graph-tool` graph generator requires a sequence of vertex degrees, describing the graph's degree distribution, to be passed to its generating function. In the graph-tool, a single sequence was used for both the in-degrees and out-degrees of the graph's vertices, essentially ensuring that the minimum total degree of a vertex is equal to twice the smallest in-degree. In the web-graph-like graphs, a total degree sequence is generated, and its values are then randomly split between the in-degree and out-degree sequences, thus allowing the generator to output vertices with a total degree of 1.
    \item Truncation of degree distribution: the graph challenge graphs truncate the degree distribution to between 10 and 100 (which effectively becomes 20 and 200 when coupled with the degree sequence duplication described above). The web-graph-like graphs have a much wider degree distribution, with the minimum degree set to 1 and the maximum degree set to a fraction of the number of vertices in the graph.
\end{itemize}

We then generate a set of 16 synthetic graphs that form an exhaustive parameter search study of these differences. These graphs are described in table~\ref{tab:ablationgraphs}. For reference, the Graph Challenge datasets are closest in structure to the TTT33 graph, while the web-graph-like graphs are closest in structure to the FFF150 graph.

\begin{table*}[htbp!]
\caption{Synthetic Graphs Used in Exhaustive Parameter Search Study}
\begin{center}
\begin{tabular}{|c|c|c|c|r||r|r|}
\hline
ID & \multirow{2}{0.1\textwidth}{\centering Truncated Min Degree} & \multirow{2}{0.1\textwidth}{\centering Truncated Max Degree} & \multirow{2}{0.15\textwidth}{\centering Duplicated Degree Sequence} & \multirow{2}{0.1\textwidth}{\centering Number of Communities} & \multirow{2}{0.1\textwidth}{\centering Number of Vertices} & \multirow{2}{0.1\textwidth}{\centering Number of Edges} \\
 & & & & & & \\
\hline
TTT33 & True & True & True & 33 & 22599 & 899283\\
TTT150 & True & True & True & 150 & 22599 & 826861 \\
TTF33 & True & True & False & 33 & 22599 & 452232 \\
TTF150 & True & True & False & 150 & 22599 & 421317 \\
TFT33 & True & False & True & 33 & 22599 & 1059970 \\
TFT150 & True & False & True & 150 & 22599 & 912644 \\
TFF33 & True & False & False & 33 & 22599 & 540410 \\
TFF150 & True & False & False & 150 & 22598 & 471071 \\
FTT33 & False & True & True & 33 & 21896 & 79683 \\
FTT150 & False & True & True & 150 & 22036 & 78226 \\
FTF33 & False & True & False & 33 & 19220 & 39719 \\
FTF150 & False & True & False & 150 & 19221 & 38408 \\
FFT33 & False & False & True & 33 & 22157 & 83939 \\
FFT150 & False & False & True & 150 & 21958 & 81298 \\
FFF33 & False & False & False & 33 & 19516 & 41378 \\
FFF150 & False & False & False & 150 & 19358 & 40835 \\
\hline
\end{tabular}
\label{tab:ablationgraphs}
\end{center}
\end{table*}

We also generate large, synthetic graph datasets for testing the scalability of distributed community detection. These datasets, which range from 11 to 300 million edges, are summarized in Table~\ref{tab:syntheticscaling}.

\begin{table}[H] 
\caption{Synthetic Graphs Used in Scaling Study}
\begin{center}
\begin{tabular}{|c|r|r|r|}
\hline
ID & \multirow{2}{0.1\textwidth}{\centering Number of Communities} & \multirow{2}{0.1\textwidth}{\centering Number of Vertices} & \multirow{2}{0.1\textwidth}{\centering Number of Edges} \\
 & & & \\
\hline
1M & 1075 & 1051218 & 11056834 \\
2M & 1521 & 2103554 & 23987218 \\
4M & 2151 & 4221264 & 53175026 \\
\hline
\end{tabular}
\label{tab:syntheticscaling}
\end{center}
\end{table}

To ensure the rigor of our experimental evaluation of DC-SBP and EDiSt, we generate all synthetic graphs to have a complex community structure in accordance with the high overlap, high block size variation graphs used in the Graph Challenge~\cite{Kao2017StreamingPartition}. That is, the ratio of intra-community to inter-community edges is roughly 2, and the approximate sizes of the communities are drawn from a random dirichlet distribution with $\alpha = 2$.

\subsection{Real-world datasets}\label{sec:realworlddata}

We also test our approach on real-world data from a variety of domains. The five graphs outlined in Table~\ref{tab:realworldscaling} are graphs from the Stanford Large Network Dataset Collection~\cite{Leskovec2014SNAPCollection}, obtained in Matrix Market format via the SuiteSparse Matrix Collection~\cite{SuiteSparseCollection}. These graphs range from 3.4 to 69 million edges in size, and do not have robust, non-overlapping ground truth communities.

\begin{table}[H]
\caption{Real-World Graphs Used in Scaling Study}
\begin{center}
\begin{tabulary}{\columnwidth}{|C|L|R|R|} 
\hline
Graph ID & Description & Number of Vertices & Number of Edges \\
\hline
Amazon & Amazon co-purchasing graph & 403394 & 3387388 \\
Patents & Citation graph in US patents & 456626 & 3774768 \\
BerkStan & Web graph containing hyperlinks & 685230 & 7600595 \\
Twitter & Twitter social network graph & 456626 & 14855842 \\
LiveJournal & LiveJournal social network graph & 4847571 & 68993773 \\
\hline
\end{tabulary}
\label{tab:realworldscaling}
\end{center}
\end{table}

\subsection{Hardware}

All our strong scaling experiments are run on the base compute nodes of the Virginia Tech tinkercliffs cluster. Tinkercliffs contains 308 base compute nodes, each equipped with 128-core AMD EPYC 7702 chips and having 256 GB of memory. The 128 cores are arranged in 8 NUMA nodes with 16 cores each, and the interconnect used is HDR-100 IB. Though the total number of base compute nodes is 308, the maximum number of nodes allowed in any slurm job is 64. Hence, our experiments are limited to 64 compute nodes.

Our smaller experiments were run on the Dell nodes of the Virginia Tech infer cluster. Infer contains 40 Dell nodes with 28 cores per node, 512Gb of memory each, and with an ethernet interconnect.

\section{Results}

In this section, we present the results of our evaluation of DC-SBP and SBP. We show that our C++ DC-SBP implementation is just as accurate as the original python implementation while being faster. We illustrate the two failure modes of DC-SBP, and show that EDiSt does not exhibit those same failure modes. Finally, we show that by leveraging its improved scalability, EDiSt provides faster runtimes and higher accuracy than DC-SBP on both synthetic and real-world graphs.

\subsection{Accuracy of our C++ DC-SBP Implementation}

To ascertain that our C++ DC-SBP implementation is just as accurate as the original python implementation, we compare the NMI results obtained at 8 MPI ranks with the original python DC-SBP implementation and our C++ DC-SBP implementation on the Graph Challenge graphs (similar to the ones for which results were reported in~\cite{Uppal2017ScalablePartition}) on the Infer cluster.
The results, summarized in Table~\ref{tab:pythonvscpp} show that the NMI obtained with our C++ implementation matches or exceeds the NMI obtained with the python implementation on all 3 graphs.

\begin{table}[htbp]
\caption{Comparison Between Python and C++ DC-SBP Implementations}
\begin{center}
\begin{tabular}{|c|r|r|r|r|}
\hline
Graph & \multicolumn{2}{c|}{Python} & \multicolumn{2}{c|}{C++} \\
ID & NMI & Runtime (s) & NMI & Runtime (s) \\
\hline
20k-easy & 0.98 & 171 & 1.00 & 26 \\
20k-hard & 0.82 & 163 & 0.86 & 30 \\
50k-easy & 0.96 & 441 & 1.00 & 100 \\
50k-hard & 0.72 & 379 & 0.84 & 178 \\
200k-easy & 0.93 & 7641 & 1.00 & 604 \\
200k-hard & 0.81 & 7244 & 0.81 & 479 \\
\hline
\end{tabular}
\label{tab:pythonvscpp}
\end{center}
\end{table}

The improved runtime performance is due to our optimizations and the use of a compiled language. The slight improvement in NMI is most likely due to our use of the Hybrid SBP algorithm for parallelizing MCMC movements within a node, compared to the batch-based parallelism employed in the original python implementation.

\subsection{Exhaustive Parameter Search Study with DC-SBP}~\label{sec:dcsbpablationresults}

We run DC-SBP with varying numbers of compute nodes on the exhaustive parameter search study graphs described in~\ref{tab:ablationgraphs} and record the resulting NMI. The results are summarized in Table~\ref{tab:dcsbpablationresults}.


These results illustrate the two conditions where DC-SBP suffers from convergence issues. The first condition occurs when the number of compute nodes (or MPI tasks) is greater than or equal to 16. The second condition occurs when the degree distribution is \textit{not} truncated on the minimum degree side (i.e., the minimum vertex degree is 1 or 2). This is because this parameter has the biggest effect on the density of the graph, and the resulting graphs are significantly more sparse than the graphs with a truncated degree distribution. This suggests that DC-SBP works much better on denser graphs, and could explain why DC-SBP maintains NMI on TFT33, the densest of these graphs, at 16 compute nodes.

\begin{table}[htbp]
\caption{NMI with DC-SBP on exhaustive parameter search graphs}
\begin{center}
\begin{tabular}{|c||r|r|r|r|r|r|r|}
\hline
Graph & \multicolumn{7}{|c|}{NMI at Number of Nodes} \\
ID & Baseline (1) & 2 & 4 & 8 & 16 & 32 & 64 \\
\hline
TTT33 & 0.92 & 0.96 & 0.89 & 0.90 & 0.87 & \cellcolor{red!25}0.00 & \cellcolor{red!25}0.00 \\
TTT150 & 0.97 & 0.97 & 0.97 & 0.96 & 0.88 & \cellcolor{red!25}0.00 & \cellcolor{red!25}0.00 \\
TTF33 & 0.96 & 0.89 & 0.95 & 0.95 & \cellcolor{red!25}0.00 & \cellcolor{red!25}0.00 & \cellcolor{red!25}0.00 \\
TTF150 & 0.95 & 0.96 & 0.96 & 0.93 & 0.67 & \cellcolor{red!25}0.00 & \cellcolor{red!25}0.00 \\
TFT33 & 0.97 & 0.97 & 0.97 & 0.91 & 0.96 & 0.73 & \cellcolor{red!25}0.00 \\
TFT150 & 0.97 & 0.97 & 0.97 & 0.96 & 0.80 & \cellcolor{red!25}0.00 & \cellcolor{red!25}0.00 \\
TFF33 & 0.97 & 0.92 & 0.89 & 0.97 & 0.65 & \cellcolor{red!25}0.00 & \cellcolor{red!25}0.00 \\
TFF150 & 0.96 & 0.96 & 0.96 & 0.93 & \cellcolor{red!25}0.00 & \cellcolor{red!25}0.00 & \cellcolor{red!25}0.00 \\
FTT33 & 0.66 & 0.53 & \cellcolor{red!25}0.00 & \cellcolor{red!25}0.00 & \cellcolor{red!25}0.00 & \cellcolor{red!25}0.00 & \cellcolor{red!25}0.00 \\
FTT150 & 0.72 & 0.59 & \cellcolor{red!25}0.00 & \cellcolor{red!25}0.00 & \cellcolor{red!25}0.00 & \cellcolor{red!25}0.00 & \cellcolor{red!25}0.00 \\
FTF33 & 0.38 & \cellcolor{red!25}0.00 & \cellcolor{red!25}0.00 & \cellcolor{red!25}0.00 & \cellcolor{red!25}0.00 & \cellcolor{red!25}0.00 & \cellcolor{red!25}0.00 \\
FTF150 & 0.48 & \cellcolor{red!25}0.00 & \cellcolor{red!25}0.00 & \cellcolor{red!25}0.00 & \cellcolor{red!25}0.00 & \cellcolor{red!25}0.00 & \cellcolor{red!25}0.00 \\
FFT33 & 0.74 & 0.66 & \cellcolor{red!25}0.00 & \cellcolor{red!25}0.00 & \cellcolor{red!25}0.00 & \cellcolor{red!25}0.00 & \cellcolor{red!25}0.00 \\
FFT150 & 0.72 & 0.69 & \cellcolor{red!25}0.00 & \cellcolor{red!25}0.00 & \cellcolor{red!25}0.00 & \cellcolor{red!25}0.00 & \cellcolor{red!25}0.00 \\
FFF33 & 0.34 & \cellcolor{red!25}0.00 & \cellcolor{red!25}0.00 & \cellcolor{red!25}0.00 & \cellcolor{red!25}0.00 & \cellcolor{red!25}0.00 & \cellcolor{red!25}0.00 \\
FFF150 & 0.48 & \cellcolor{red!25}0.00 & \cellcolor{red!25}0.00 & \cellcolor{red!25}0.00 & \cellcolor{red!25}0.00 & \cellcolor{red!25}0.00 & \cellcolor{red!25}0.00 \\
\hline
\end{tabular}
\label{tab:dcsbpablationresults}
\end{center}
\end{table}

\begin{figure}[H]
  \centering
  \includegraphics[width=0.8\columnwidth]{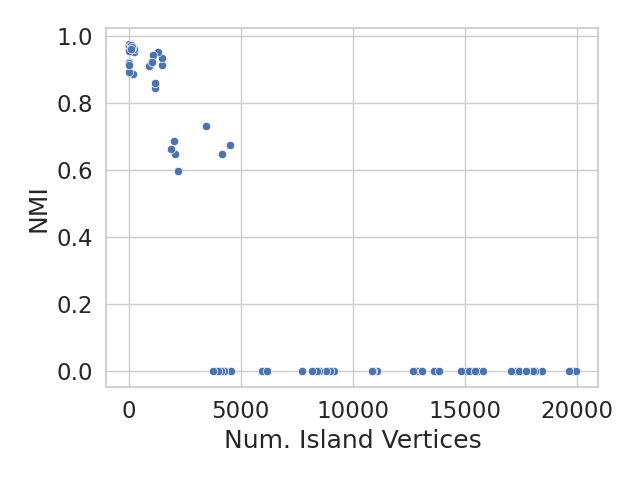}
  \caption{Relationship between the number of island vertices induced by the data distribution and the resulting NMI.}
  \label{fig:nmivsislands}
\end{figure}

We attribute this to the number of island vertices induced by the data distribution method in DC-SBP. This is corroborated by Figure~\ref{fig:nmivsislands}, which shows the relationship between the number of island vertices and the NMI obtained on the results in Table~\ref{tab:dcsbpablationresults} above. As the number of island vertices increases, the NMI decreases. DC-SBP appears to be robust up to a relatively high number of island vertices of around 10\%, but NMI drops off after that and rests at 0 beyond 20\%. Tellingly, on the sparser graphs, and with a high number of compute nodes, the data distribution scheme results in upwards of 50\% island vertices.

\subsection{Exhaustive Parameter Search Study with EDiSt}

We repeat the exhaustive parameter search study using EDiSt on the same set of graphs and record the results in Table~\ref{tab:EDiStablationresults}. Our results demonstrate that EDiSt converges well even when the number of compute nodes is high, and the graphs are sparse due to a power law degree distribution with the minimum degree being 1.

\begin{table}[htbp]
\caption{NMI with EDiSt on exhaustive parameter search graphs}
\begin{center}
\begin{tabular}{|c||r|r|r|r|r|r|r|}
\hline
Graph & \multicolumn{7}{|c|}{NMI at Number of Nodes} \\
ID & Baseline (1) & 2 & 4 & 8 & 16 & 32 & 64 \\
\hline
TTT33 & 0.92 & 0.88 & 0.94 & 0.93 & 0.95 & 0.95 & 0.94 \\
TTT150 & 0.97 & 0.97 & 0.97 & 0.97 & 0.97 & 0.97 & 0.97 \\
TTF33 & 0.96 & 0.89 & 0.96 & 0.96 & 0.95 & 0.96 & 0.96 \\
TTF150 & 0.95 & 0.95 & 0.95 & 0.96 & 0.97 & 0.97 & 0.95 \\
TFT33 & 0.97 & 0.96 & 0.98 & 0.98 & 0.97 & 0.96 & 0.96 \\
TFT150 & 0.97 & 0.97 & 0.97 & 0.97 & 0.97 & 0.97 & 0.97 \\
TFF33 & 0.97 & 0.96 & 0.96 & 0.97 & 0.97 & 0.96 & 0.96 \\
TFF150 & 0.96 & 0.95 & 0.97 & 0.96 & 0.95 & 0.95 & 0.95 \\
FTT33 & 0.66 & 0.66 & 0.66 & 0.67 & 0.67 & 0.65 & 0.66 \\
FTT150 & 0.72 & 0.72 & 0.72 & 0.73 & 0.72 & 0.70 & 0.72 \\
FTF33 & 0.38 & 0.38 & 0.38 & 0.35 & 0.36 & 0.34 & 0.38 \\
FTF150 & 0.48 & 0.42 & 0.48 & 0.44 & 0.49 & 0.39 & 0.48 \\
FFT33 & 0.74 & 0.75 & 0.74 & 0.74 & 0.75 & 0.75 & 0.73 \\
FFT150 & 0.72 & 0.71 & 0.72 & 0.72 & 0.72 & 0.75 & 0.72 \\
FFF33 & 0.34 & 0.35 & 0.38 & 0.36 & 0.38 & 0.38 & 0.40 \\
FFF150 & 0.48 & 0.51 & 0.53 & 0.51 & 0.51 & 0.52 & 0.53 \\
\hline
\end{tabular}
\label{tab:EDiStablationresults}
\end{center}
\end{table}

\subsection{Scalability Analysis on Synthetic Graphs}

In this subsection, we discuss the scalability of EDiSt on synthetic graphs, and compare the results to those obtained with DC-SBP.

\subsubsection{Scalability on a Single Node}

The hybrid MCMC method described in~\cite{Wanye2022OnDetection} alternates between sequential and parallel MCMC execution. This leads to extended periods of low CPU utilization where only one thread is running. In such cases, running EDiSt on multiple MPI tasks on the same node can improve SBP runtime, provided the node has enough memory and the total number of tasks is significantly smaller than the number of high-degree vertices in the graph (in this case, the hybrid parallel MCMC algorithm would devolve into asynchronous Gibbs, which could reduce accuracy on certain graphs).

\begin{figure}[htbp]
  \centering
  \includegraphics[width=0.7\columnwidth]{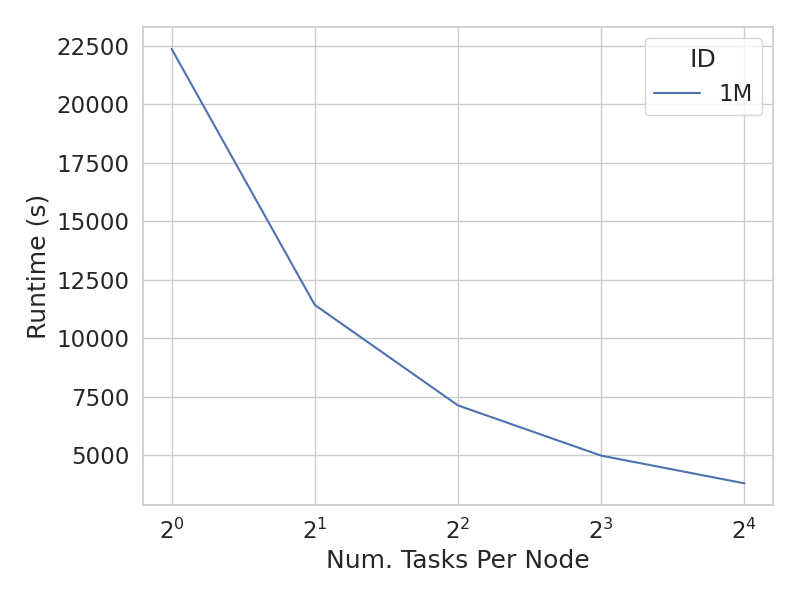}
  \caption{EDiSt runtime with multiple MPI tasks per compute node.}
  \label{fig:singlenodescalability}
\end{figure}

In Figure~\ref{fig:singlenodescalability}, we showcase this increase in speedup by running EDiSt on the 1M graph described in Table~\ref{tab:syntheticscaling} on a single node, and increasing the number of MPI tasks until we run out of memory.

Our results indicate that there is a runtime benefit to running multiple EDiSt MPI tasks per node, with a speedup of 9$\times$ with 16 MPI tasks per node. In all proceeding sections, we run EDiSt with 4 MPI tasks per node (we run out of memory when attempting more tasks per node with larger graph sizes). We do not do the same for DC-SBP, because the convergence issues observed when the number of MPI ranks is increased make such a solution impractical for that algorithm.

\subsubsection{Strong Scaling}

\begin{figure}[htbp]
  \centering
  \includegraphics[width=0.7\columnwidth]{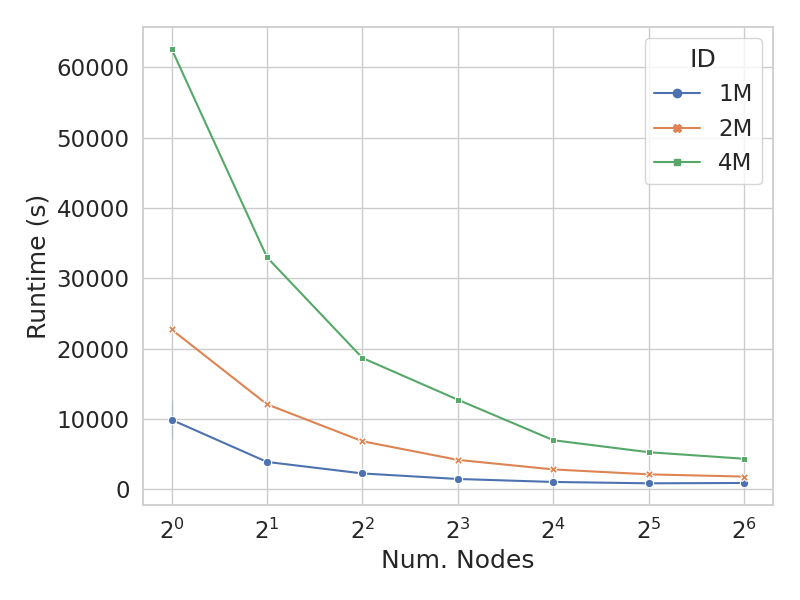}
  \includegraphics[width=0.7\columnwidth]{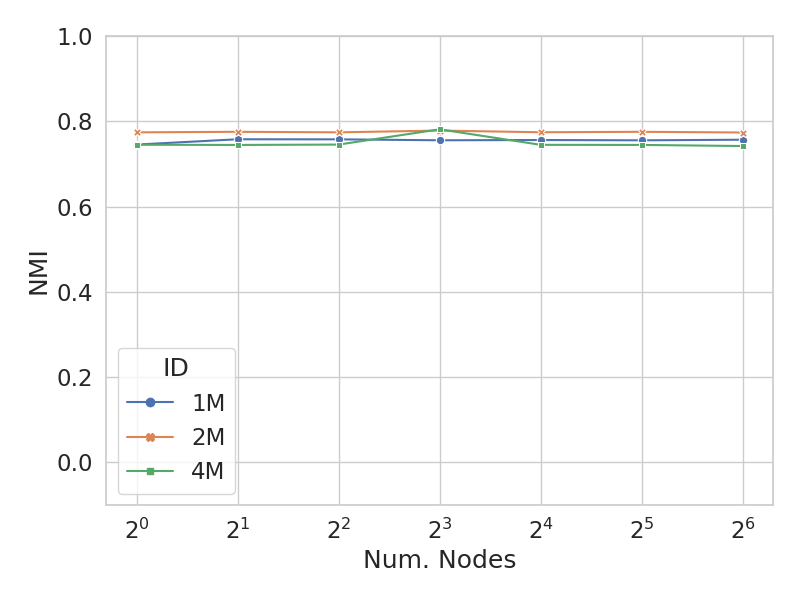}
  \caption{EDiSt strong scaling (top) and NMI (bottom) results on synthetic graphs.}
  \label{fig:EDiStscaling}
\end{figure}

We run EDiSt on the synthetic scaling graphs we describe in Table~\ref{tab:syntheticscaling}, and measure the resulting NMI and Runtime. Figure~\ref{fig:EDiStscaling} shows the runtime and NMI results of EDiSt as the number of nodes increases from 1 to 64.

These results confirm that EDiSt maintains result quality both at high numbers of MPI tasks and on sparse graphs. Given that we run EDiSt with 4 MPI tasks per node, EDiSt is therefore usable on at least 16$\times$ more MPI tasks than DC-SBP. Though the runtimes do start to level off, the level off point increases as the graph size increases, suggesting that up to 64 nodes, the runtime benefits of EDiSt will scale with the graph size.

\subsubsection{Comparison with DC-SBP}

\begin{figure}[H]
  \centering
  \includegraphics[width=1.0\columnwidth]{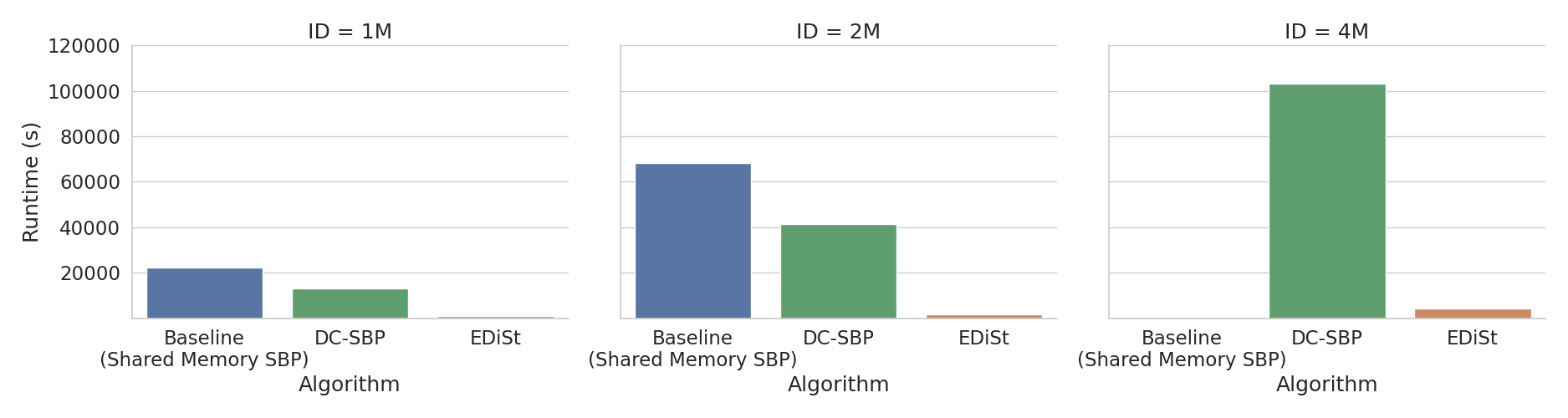}
  \caption{Comparison of the best runtimes achieved with DC-SBP and EDiSt on the three synthetic scaling graphs.}
  \label{fig:syntheticcomparison}
\end{figure}

We then compare the speedups obtained with EDiSt to those obtained with DC-SBP. In the case of DC-SBP, for each graph we select the runtime at the highest number of MPI tasks at which DC-SBP maintains NMI with the single node shared-memory baseline. The results are shown in Figure~\ref{fig:syntheticcomparison}. Results were omitted wherever the runtime exceeds 36 hours.

These results show that EDiSt on 64 compute nodes is up to 38.0$\times$ faster than single node shared memory SBP, and up to 23.8$\times$ faster than the best-performing DC-SBP run that did not suffer from convergence issues.

\subsection{Real-World Graphs Results}

\begin{figure*}[htbp!]
  \centering
  \includegraphics[width=0.9\textwidth]{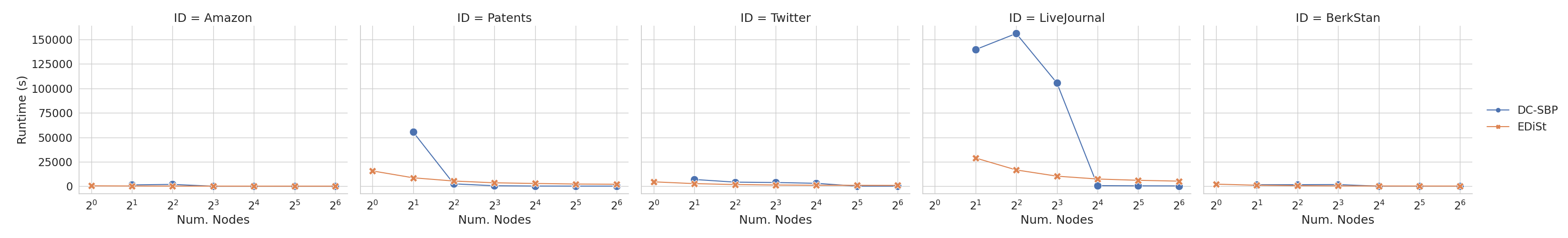}
  \includegraphics[width=0.9\textwidth]{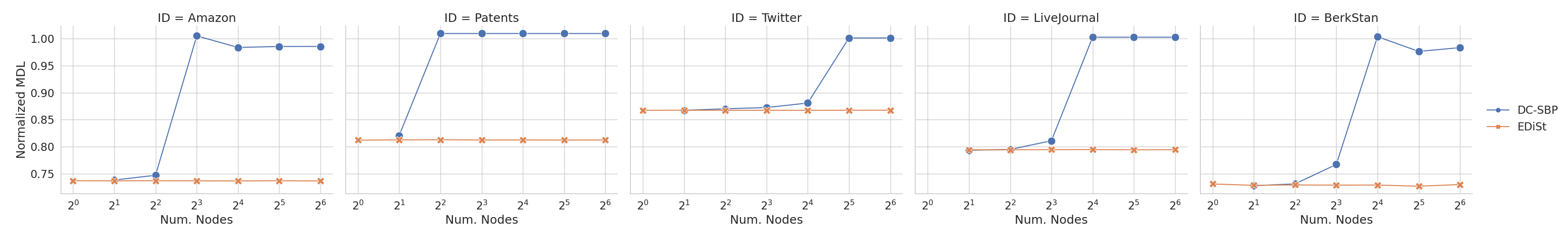}
  \caption{EDiSt vs DC-SBP strong scaling (top) and accuracy as measured by Normalized MDL (bottom, lower is better) results on real-world graphs.} 
  \label{fig:realworldresults} 
\end{figure*}

Finally, we run both DC-SBP and EDiSt on the real-world datasets described in Table\ref{tab:realworldscaling}. Because real-world graphs do not have known ground truth communities, we measure accuracy using the normalized description length~($DL^{norm}$) metric~\cite{Wanye2022OnDetection}, which is given by the equation: $DL^{norm} = \frac{DL}{DL^{null}}$, where $DL$ refers to the description length of the blockmodel returned by either algorithm, and $DL^{null}$ refers to the description length of a null blockmodel where all vertices are assigned to the same singular community.

The $DL^{norm}$ and strong scaling results obtained are summarized in Figure~\ref{fig:realworldresults}. EDiSt is up to 26.8$\times$ faster than DC-SBP on these graphs, because DC-SBP usually produces meaningful results on only 2-8 subgraphs. On the Twitter graph, which has the highest average degree, DC-SBP scales well up to 16 subgraphs, and EDiSt is only 3.4$\times$ faster than DC-SBP. Therefore, these results corroborate our findings on synthetic graphs. 


\subsection{Discussion}

Our results show that DC-SBP has convergence issues on sufficiently sparse graphs, and when the number of compute nodes~(and therefore, distinct subgraphs) is 16 or higher. Thus, it's applicability is limited to dense graphs on small clusters. On the other hand, the exact distributed stochastic block partitioning~(EDiSt) algorithm converges in both of these scenarios, making it applicable in a much wider variety of situations. In addition to this, the single-node partial result combination in DC-SBP leads to severe bottlenecks on larger graphs with more communities, to the point where the theoretically slower EDiSt outperforms it on the same number of nodes.

However, DC-SBP does have one major advantage over EDiSt - it incorporates data distribution. With the advent of web-scale graphs, which contain on the order of billions of edges, data distribution is becoming more and more important for large graph processing. However, given the current state of the SBP algorithm, we argue that runtime is a bigger bottleneck than memory usage. For example, the 69 million edge LiveJournal graph takes over an hour to process on 64 nodes. More importantly, we were able to load in a 300M edge synthetic graph on our cluster, but 
could not complete its processing on the tinkercliffs cluster within 8 hours on 64 nodes.
Additionally, data reduction techniques like sampling, which have been shown to preserve community structure in graphs~\cite{Wanye2019FastSampling,Stanley2018CompressingNodes,Maiya2010SamplingStructure}, are a promising means of reducing the memory footprint of graphs that do not fit in memory.

\section{Conclusion}

Community detection is an important graph analytics task with applications in a wide variety of fields, from bioinformatics to social media analytics. Due to the growth in size of modern graph datasets, sequential community detection algorithms are no longer practical for many real-world scenarios. This presents a problem for relatively slow and hard-to-parallelize algorithms like stochastic block partitioning~(SBP), which has largely limited its applicability in prior work to smaller datasets. In this work, we take a significant step towards making SBP practical on web-scale graphs by distributing its computation on multiple nodes of a computational cluster.

We first empirically show that the state-of-the-art distributed SBP algorithm, the divide-and-conquer SBP~(DC-SBP) algorithm, has two conditions under which it suffers from poor convergence. The first occurs when the number of compute nodes is high, and the second occurs when the graph is sufficiently sparse. In both cases, the quality of community detection results is negatively affected, often to the point where the algorithm fails to converge entirely. This effect on convergence is largely a result of the combination of the round-robin distribution strategy and lack of inter-node communication within DC-SBP, which lead to subgraphs being independently processed with many island vertices.

We then introduce our exact distributed SBP~(EDiSt) algorithm, which tackles both conditions by a) allowing data to be duplicated across MPI tasks, and b) allowing communication between the subgraphs being processed. We empirically show that EDiSt maintains result quality both at large numbers of MPI tasks, and on sparse graphs.

Technically speaking, DC-SBP is close to being computationally optimal, and thus should be faster than EDiSt given the same number of MPI tasks. However, the fact that the DC-SBP partial subgraph results are combined on just a single node presents a significant bottleneck.
Additionally, EDiSt can scale to a much larger number of MPI tasks without sacrificing result quality. Due to these two factors, EDiSt is faster than DC-SBP when run on computational clusters.
In our results, using 64 compute nodes and 256 MPI tasks, we achieve speedups as high as 26.9$\times$ over the best-performing DC-SBP run on the same real-world graph, and as high as 23.8$\times$ on the same synthetic graph. Additionally, EDiSt is up to 38.0$\times$ faster than shared memory parallel SBP on a single node.

The two approaches are not antithetical.
Unlike DC-SBP, EDiSt lacks a proper data distribution method. Implementing a distributed data structure for EDiSt is untrivial due to the need for both column-wise and row-wise traversal of the underlying blockmodel matrix, as well as the random memory access pattern that is a result of the randomness inherent to SBP.

Therefore, to incorporate data distribution into EDiSt, in future work we will be looking to improve DC-SBP performance on sparse graphs by altering its data distribution and partial result combination strategies, and then processing each subgraph using EDiSt.
Such a hybrid method could benefit from both the data distribution of DC-SBP and the scalability of EDiSt. 
We would also like to scale EDiSt to larger 
clusters. With a large number of nodes, the all-to-all communication patterns in EDIST are likely to present a significant bottleneck. To mitigate this, we plan to explore alternative communication approaches, including MPI one-sided communication primitives and prioritizing communication based on the Fisher information content of the edges in the graph~\cite{Kao2019HybridInteractions}.

\section*{Acknowledgment}

This project was supported in part by NSF I/UCRC CNS-1822080 via the NSF Center for Space, High-performance, and Resilient Computing (SHREC).

The authors acknowledge Advanced Research Computing at Virginia Tech for providing computational resources and technical support that have contributed to the results reported within this paper. URL: https://arc.vt.edu/

\bibliographystyle{IEEEtran}
\bibliography{references}

\end{document}